\documentclass[
]{template/ceurart}

\usepackage{listings}
\usepackage{svg}
\usepackage{amsmath}
\newcommand{\qoracle}{\textsf{QOracle}}

\begin{document}

  \copyrightyear{2025}
  \copyrightclause{Copyright for this paper by its authors. Use permitted under Creative Commons License Attribution 4.0 International (CC BY 4.0).}

  \conference{}

\title{Document Quality Scoring for Web Crawling}

\author[1]{Francesca Pezzuti}
\address[1]{University of Pisa, Pisa, Italy}

\author[2]{Ariane Mueller}
\address[2]{University of Glasgow, Glasgow, United Kingdom}

\author[2]{Sean MacAvaney}

\author[1]{Nicola Tonellotto}

\begin{abstract}
    The internet contains large amounts of low-quality content, yet users expect web search engines to deliver high-quality, relevant results. The abundant presence of low-quality pages can negatively impact retrieval and crawling processes by wasting resources on these documents.
    Therefore, search engines can greatly benefit from techniques that leverage efficient quality estimation methods to mitigate these negative impacts.
    Quality scoring methods for web pages are useful for many processes typical for web search systems, including static index pruning, index tiering, and crawling.
    Building on work by Chang et al.~\cite{chang2024neural}, who proposed using neural estimators of semantic quality for static index pruning, we extend their approach and apply their neural quality scorers to assess the semantic quality of web pages in crawling prioritisation tasks. 
     In our experimental analysis, we found that prioritising semantically high-quality pages over low-quality ones can improve downstream search effectiveness.
     Our software contribution consists of a Docker container that computes an effective quality score for a given web page, allowing the quality scorer to be easily included and used in other components of web search systems.
\end{abstract}


\maketitle

\section{Introduction}
The Internet contains vast amounts of information, yet not all of it is of high quality. In fact, the web is filled with low-quality web pages, including meaningless pages, keyword-stuffed pages, and spam~\cite{DBLP:journals/ir/CormackSC11}. If not properly managed, this abundance of low-quality pages, can pose significant challenges for web search engines, whose primary goal is to deliver high-quality, relevant results to users. Firstly, low-quality pages may introduce unnecessary overheads: they need to be crawled, indexed, and processed at query time, consuming resources that could be better spent on high-quality web pages and ultimately slowing down search systems. Secondly, despite being semantically poor, low-quality pages may still rank highly in search results (especially from systems that do not consider semantic quality, such as lexical retrievers), negatively affecting retrieval effectiveness. To improve efficiency and effectiveness of web search systems, a heuristic for estimating page quality is thus highly valuable. With such a heuristic, search engines can mitigate the negative impacts of low-quality pages.
For instance, one technique that can be employed to speed up \textit{indexing and ranking} times while decreasing memory requirements, is \textit{static index pruning}. This approach discards low-quality pages from the search index.
As an example of this, Chang et al. recently proposed a neural quality estimator that approximates semantic quality and has proven strongly effective for static indexing pruning~\cite{chang2024neural}.
Meanwhile, to speed up the retrieval of high-quality results \textit{at query processing time}, search engines can implement \textit{tiered indexing}~\cite{DBLP:conf/sigir/Baeza-YatesMH09}. This approach consists in organising the index in multiple tiers based on page quality: high-quality pages are placed in the top tier to be retrieved quickly, low quality-ones are placed in lower tiers and are processed only when necessary.
Finally, quality estimation is crucial in the \textit{crawling stage} of a web search engine, where the goal is to traverse the web link graph and download pages to build a corpus of documents. Specifically, the crawler component of a search systems often uses quality estimation heuristics to prioritise the download of high-quality web pages over low-quality ones, aiming to improve the early downstream ranking effectiveness of the retriever component.
Common quality estimation methods for crawling prioritisation include those based on connectivity metrics like PageRank~\cite{page1999pagerank} and indegree~\cite{kumar2008indegree}, or Click-Through-Rate~\cite{ostroumova2014ctr}. However, these quality estimation techniques often have high computational demands~\cite{menczer2004topiccrawl}, require storing information about the web graph, or the previous popularity of web pages~\cite{ostroumova2014ctr}. More importantly, to the best of our knowledge, existing prioritisation techniques for crawlers do not account for the \textit{semantic quality} of web pages.

Considering the semantic quality of pages during crawling has substantial potential. \citet{chang2024neural} showed that these signals can be useful for static pruning, and \citet{yu2025craw4llm} showed that they can be useful for identifying language model pre-training data. Therefore, in this work, we test whether semantic quality signals are helpful for prioritising web pages to crawl for a search engine. The central hypothesis is that documents of similar semantic quality will likely link to one another: high-quality to high-quality and low-quality to low-quality.\footnote{This intuition matches the intuition of PageRank~\cite{page1999pagerank}, but considers the semantic quality of the document contents, rather than the link structure.} By leveraging this signal, we anticipate that we can both identify high-quality pages faster during crawling and avoid wasting resources on low-quality pages. To assess this goal, we implement a dockerised quality scoring module. As this approach is containerised, it can be easily included and used in Open Web Search (OWS) components~\cite{ows-book} to compute document quality scores. Additionally, we integrate this quality scoring approach within the Resilipipe pre-processing pipeline~\cite{heineking_2024_13784624}.
Using our implementation, we score subsets of the main and legal collections of OWS datasets~\cite{owlerdashboard}. 
However, since these two collections are not associated with a web graph, we conduct a proof of concept on crawling prioritisation strategies based on neural quality estimators using the English subset of the ClueWeb22-B~\cite{overwijk2022clueweb22} web corpus. Our analysis reveals that the distribution of quality scores of the two OWS datasets closely matches that of the English subset of ClueWeb22-B, suggesting that our findings for this dataset are likely to generalise to OWS.

Specifically, our preliminary findings on ClueWeb22-B~\cite{overwijk2022clueweb22} show that by prioritising web pages with high semantic quality over those with lower quality, relevant content is implicitly prioritised over irrelevant content. Our experiments show that an oracle crawler leveraging a semantic quality scorer improves early downstream recall effectiveness compared to two well-known graph-traversal crawling strategies, namely Breadth-First-Search~\cite{najork2001breadth} and Depth-First-Search crawlers~\cite{debra1994fishsearch}. Furthermore, because the quality of a page is positively correlated with that of its (outlinking) neighbours, its quality score can serve as an estimate for the quality of pages it links to. Consequently, in real-world crawling scenarios where the text of a web page is unavailable before its download, these estimates can effectively be used to prioritise semantically valuable web pages, to skip pages mostly linked to by low-quality pages, or to avoid crawling domains that mostly host low-quality pages.

The remainder of this paper is organised as follows. First, in Section~\ref{methods} we describe the quality scorer and web crawler used for the mentioned proof of concept experiments. In Section \ref{software} we provide details on our implementation of quality scoring for OWS datasets. Then, in Section \ref{experimental setup} we describe our experimental setup. In Section~\ref{analysis} we show and discuss results of our experiments. Lastly, in Section \ref{conclusion} we summarise our contribution and give an outlook on potential future work.

\section{Methodology}\label{methods}

A \textit{quality scorer} component for a web search system consists of a large language model (LLM) (denoted as $\mathcal{Q}_\theta$), trained to distinguish pages of high semantic quality from low-quality ones~\cite{chang2024neural}. Formally, a quality scorer parametrised by $\theta$ is characterised by a \textit{quality scoring function} $\mathcal{Q}_\theta:t\mapsto \mathbb{R}$ that estimates the semantic quality of a text $t$ with a real valued score  $q = \mathcal{Q}_\theta(t)$. This value is referred to as a \textit{quality score}.
Because quality scorers are based on LLM models, they can efficiently generate quality scores in batches, making them suitable for simultaneously scoring multiple text documents.

These quality scorers can be applied in a variety of contexts, both to improve efficiency and effectiveness. For example, they have already proven effective for static pruning tasks~\cite{chang2024neural}. Another promising application is the prioritisation of web pages characterised by high semantic quality during the crawling process, to build semantically high-quality corpora  and improving downstream effectiveness.

The \textit{crawler} component of a web search system systematically traverses the web graph following hyperlinks and downloading web pages valuable for downstream tasks like retrieval. To achieve this, it maintains a priority queue of links yet to be crawled, ordered by quality. We propose using the quality scorer described in the previous section to assign a quality score to each page, which is then used to determine its crawling priority in the queue. However, in real-world crawling scenarios the text of a page is unavailable before its crawl. Therefore, such an approach is only applicable if it is combined with an oracle function $\mathcal{O}: x\mapsto t$ that provides the text $t$ of a web page $x$ before the page is actually downloaded.
We call this approach \qoracle.
Given a quality scorer $\mathcal{Q}_\theta$ and an oracle function $\mathcal{O}(\cdot)$, the \qoracle\ crawler computes the crawling priority $p_x$ of a web page $x$ as:
\begin{equation*}
    p_x =  \mathcal{Q}_\theta \left (\mathcal{O}(x) \right) = \mathcal{Q}_\theta(t) = q_x,
\end{equation*} 
where the priority $p_x$ is a real-valued score reflecting the level of semantic quality $q_x$ of the textual content $t$ of page $x$ to be crawled.

However, in most real-world scenarios we do not have access to an oracle function. Consequently, crawlers need to rely on alternative methods based on quality approximations. Indeed, if the quality of a page reflects the quality of its neighbours in the web graph, one could approximate the quality of an un-crawled linked page using the quality of the in-linking page during crawling prioritisation.
To test this hypothesis, we measure the linear correlation between the quality of a page $x$ and the mean quality of the pages it links to.
Formally, we denote with $\mathcal{N}:x\mapsto \left \{ l_1, \ldots, l_N\right\}$ the function that given a page $x$, returns the set of pages $\mathcal{N}(x)=\left \{l_1\, \ldots, l_N\right\}$ with an incoming link from $x$. We also denote with $\hat{q}_x \in \mathbb{R}$ the mean quality of pages in $\mathcal{N}(x)$, and we compute it as:
\begin{equation*}
    \hat{q}_x = \frac{1}{\vert \mathcal{N}(x)\vert} \sum_{l \in \mathcal{N}(x)} q_l,
\end{equation*} where $q_l$ is the quality score of a neighbour page of $x$, computed with $\mathcal{Q}_\theta$.
To investigate how the quality of a page relates to the quality of the pages it links to, we propose to measure how strongly the quality of a page $q_x$ correlates to the quality $\hat{q}_x$ of its out-linking pages on a collection of web pages.

\section{Software Implementation}\label{software}
In this section we provide details on the architecture and implementation of our containerised quality scoring module as well as the integration of quality scoring within the Resilipipe preprocessing pipeline.
\subsection{Quality Scoring Module}\label{methods: container}

\begin{figure}
    \centering
    \includegraphics[width=0.8\linewidth]{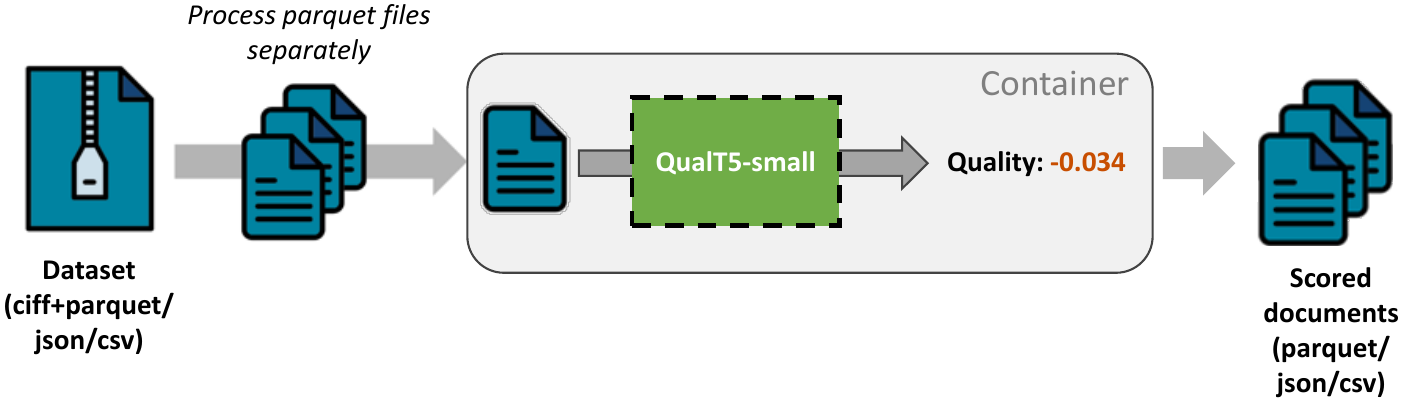}
    \caption{Architecture of our containerised quality scorer module. The quality score is added to the input as an additional column. The input and output format do not have to be identical.}
    \label{fig:container}
\end{figure}

In order to make our approach easily deployable in a crawling or pre-processing scenario, we sandbox our application within a custom Docker container as shown in Figure~\ref{fig:container}. While the standard input and output file format of the container is \texttt{parquet} (in accordance with the file format used for the pre-processed OWS datasets, i.e. datasets in owi-format~\cite{ows-book}, which are described in further detail in Section~\ref{experimental setup}), it also offers support for several other formats, including \texttt{csv} and \texttt{json}. This allows the quality scorer to be easily integrated and used in other components of the web index of OWS.

\subsection{Integration with Resilipipe}
In addition to the dockerised quality scoring module, we integrate quality scoring as an additional module of the Resilipipe preprocessing pipeline \cite{heineking_2024_13784624}. This pipeline receives the crawled \texttt{warc} files, containing the full HTTP request/response stream from the crawling process, as input and extracts information such as outlinks, plain text, geoinformation and the language of the respective website(s). As a last step, the websites/ documents are indexed and the extracted metadata is provided in \texttt{parquet} files. Adding quality scoring of documents to the pipeline allows to use the obtained scores during indexing (or other post-processing steps) e.g. for filtering out low-quality documents that should not be indexed, therefore reducing index size and retrieval latency \cite{chang2024neural}. 

\begin{figure}[htbp!]
    \centering
    \includegraphics[width=\linewidth]{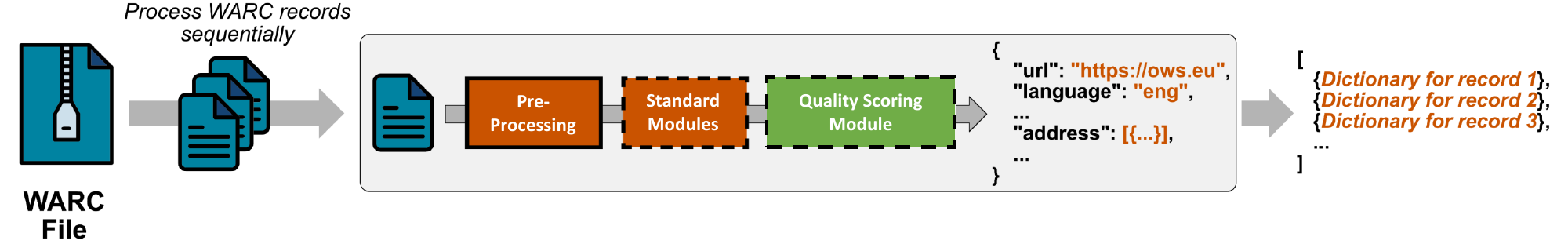}
    \caption{Overview of the integration of our quality scoring module (highlighted in green) in the Resilipipe pre-processing pipeline,   adapted from \cite{ows-book}.}
    \label{fig:pipeline}
\end{figure}

For our module, we employ the same QT5-small based scorer as used for the containerised version. The model receives the plaintext of a document, which was extracted in a previous step, and produces the corresponding quality score. The score is added as an additional column in the extracted metadata and is accessible to post-processing modules in the pipeline. The full architecture of the pipeline is visualised in Figure \ref{fig:pipeline}. The pre-processing step parses the HTML, extracting the HTML tree and the plaintext of a document, as well as metadata such as language. The standard modules are additional pre-implemented modules that extract further metadata such as link or geo-information \cite{heineking_deliverable}. After the standard and quality scoring modules, the resulting data is stored in \texttt{parquet} files and the documents are indexed.

\section{Experimental Setup}\label{experimental setup}
In this section we describe the setup used in our experimental analysis, and provide details on the used datasets and training procedures for the neural quality scorer.

\paragraph{Quality scoring for Open Web Search datasets}
We deploy the containerised quality scoring module described in Section \ref{methods: container} to estimate the document quality of several OWS datasets. As our quality estimator, we use the QT5-small based model trained by Chang et al. \cite{chang2024neural} without further fine-tuning. It is important to note that this and all other quality scoring models used in our experiments assign the \textit{log-probability} for a document of being relevant to at least one user query as its quality score. In particular, we score English documents from (arbitrarily) selected subsets of the legal and main collections of OWS~\cite{owlerdashboard}. All used datasets are in owi-format, i.e. they consist of several \texttt{parquet} files containing document (meta-)data as well as a \texttt{ciff} index file.
Table~\ref{tab: scored datasets} shows an overview of all scored datasets. It is to be noted that the OWS main and legal collections as well as our subsets of them are not mutually exclusive, i.e. their documents may overlap. However, as mentioned in the introduction, the OWS datasets do not provide a corresponding web graph nor a set of queries associated with relevance judgements. Hence, we use the ClueWeb22-B dataset for further experimentation as follows.

\begin{table}
    \centering
    \begin{tabular}{ccccc}
        \toprule
         Collection&Datacenter&Crawl Date&ID&Original size\\
         \midrule
         main&it4i&22/01/2024&03858e76-3308-11ef-a49c-0242ac1d000a&1.2 GB\\
         main&lrz&27/11/2023&25e96c3a-7477-4ebe-886a-57aa157c3425&18 GB\\
         main&lrz&23/11/2023&cd0c41ff-a5d2-493c-9e4f-249939106323&7.5 GB\\
         main&lrz&20/11/2023&93150510-14b4-49e9-96a6-4c765477dfbd&41 GB\\
         main&lrz&25/10/2023&e36b4256-4c98-4d27-b675-3b70f7d04daf&42 GB\\
         main&lrz&24/10/2023&a7c072ba-f64c-44e2-befb-0ac99edde800&1.4 GB\\
         main&lrz&16/11/2023&989a458e-16ff-43dd-bee5-97295587f7c6&38 GB\\
         main&lrz&21/11/2023&9c540793-726b-484b-b496-92f3df4e65aa&22 GB\\
         legal&it4i&12/03/2024-21/03/2024&0dac12be-52f5-11ef-a60f-0242c0a81003&6.7 GB\\
         legal&it4i&01/01/2024-31/01/2024&1459dc0c-4e42-11ef-b6de-0242c0a81003&6.9 GB\\
         legal&it4i&03/12/2023-25/12/2023&33d3b674-4e5c-11ef-8f9d-0242c0a81003&3.4 GB\\
         legal&it4i&04/02/2024-16/02/2024&af54360a-4f08-11ef-af7b-0242c0a81003&0.9 GB \\
         \bottomrule
    \end{tabular}
    \caption{Overview of the OWS datasets we selected for applying quality scoring.}
    \label{tab: scored datasets}
\end{table}

\paragraph{Quality scoring for ClueWeb22-B}
To estimate the quality for pages in the subset of $87$ million English head web pages of  ClueWeb22-B~\cite{overwijk2022clueweb22} (ClueWeb22-B (en)), and being coherent with the scoring performed on OWS datasets, we again use the QT5 quality estimator without further fine-tuning it.

\paragraph{Crawling ClueWeb22-B}
To perform the experiments on ClueWeb22-B (en), we fine-tune our QT5-small model using a version of the training procedure provided in the original paper~\cite{chang2024neural}, modified to work with the ClueWeb22 dataset. In particular, we sample $9.1$ million documents with positive relevance label from the MS MARCO Web Search dataset~\cite{chen2024msmarco}, and we use them as positive quality labels, considering the remaining set of documents as negatives. The model converged after $1.6$ million training instances. Our QT5-small quality scorer model fine-tuned on MS MARCO Web Search, is available on HuggingFace\footnote{\url{https://huggingface.co/macavaney/qt5-small-msw}}.
In our simulations of crawling processes, we always start from a fixed set of $100$ thousand randomly selected seed pages, and we reach a total of $29$ million pages. To evaluate the early downstream effectiveness of our oracle crawler, we index the crawled corpora  after every $5$ million pages have been crawled.
We compare our proposed oracle crawler using Breadth-First-Search (BFS)~\cite{najork2001breadth} and Depth-First-Search (DFS)~\cite{debra1994fishsearch} crawlers as baselines.

\paragraph{Query sets \& Downstream Retrieval effectiveness}
To evaluate crawlers in terms of downstream retrieval effectiveness, we measure the recall at cutoff $100$, of a BM25 retriever~\cite{robertson1994bm25}.
We use a mix of queries from the Researchy Questions query set (RQ)~\cite{rosset2024rqdataset} and MS MARCO Web Search query set (MSM-WS)~\cite{chen2024msmarco}. Both these two datasets are generated from the logs of commercial search engines. While MSM-WS contains explicit relevance assessments extracted from a real click-log, RQ only provides a click distribution; thus, for queries from RQ, we consider as relevant the most clicked page. 
In particular, we measure the retrieval recall at cutoff $100$ (R@100), for a query set composed of $850$ queries randomly selected from RQ, and $850$ queries randomly selected from MSM-WS.  For significance testing we conduct Bonferroni-corrected pairwise t-tests with $p\le0.01$.

\paragraph{Code, Container, and Data Availability}
We publish on Github the code for our dockerised quality scoring component\footnote{ \url{https://github.com/ArianeS21/quality_scoring}} as well as our custom Resilipipe module\footnote{ \url{https://github.com/ArianeS21/resilipipe_quality_scoring}}. We also provide the code for reproducing experiments and crawling simulations on ClueWeb22-B (en)\footnote{https://github.com/fpezzuti/quality\_crawling}.
The quality scores computed on OWS datasets are available on Zenodo\footnote{\url{https://zenodo.org/records/15110099}}~\cite{mueller2025qscore}, whereas those computed on ClueWeb22-B (en) are available on HuggingFace\footnote{macavaney/cw22b-en.qt5-small-msw.cache}.

\section{Experimental Analysis}\label{analysis}
In this section we describe and discuss our experiments and their results.

\subsection{Comparison between quality score distributions}

\begin{figure}
    \centering
    \includegraphics[width=\textwidth]{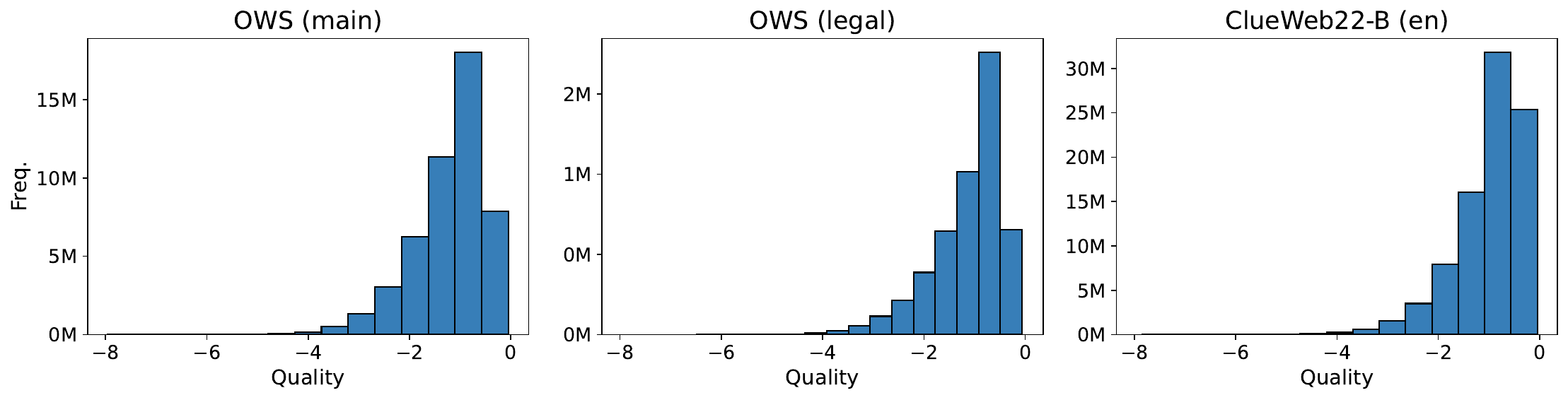}   
    \caption{Comparison between the distributions of the quality score computed for subsets of OWS (main, legal), and for ClueWeb22-B. All the histograms are generated using $15$ bins.}
    \label{fig:hists_comparison}
\end{figure}

To gain an insight into the similarity between the OWS collections and ClueWeb22-B w.r.t. document quality, we measure and compare quality scores computed on the three datasets described in Section \ref{experimental setup}, i.e., OWS (main), OWS (legal), and ClueWeb22-B (en). Figure~\ref{fig:hists_comparison} shows the distributions of quality scores for the respective datasets.
Based on the shown histograms, we note that the three considered datasets exhibit very similar quality distributions, especially the two OWS collections. 
This result is also quantitatively confirmed by the values of the Jensen-Shannon distance values computed for pairs of histogram distributions, as shown in Table~\ref{tab:jsdistance}.
\begin{table}[]
    \centering
    \begin{tabular}{lccc}
        \toprule
          & OWS (main) & OWS (legal) & ClueWeb22-B (en)\\
          \midrule
         OWS (main) & $0$  & - & - \\
         OWS (legal) & $0.0927$ & $0$ & - \\
         ClueWeb22-B (en) & $0.1437$ & $0.2015$ & $0$ \\
         \bottomrule
    \end{tabular}
    \caption{Jensen-Shannon distance between pairs of histogram distributions of quality scores computed for subsets of OWS (main, legal) and for ClueWeb22-B (en) using $15$ bins.}
    \label{tab:jsdistance}
\end{table}
The OWS legal collection is a subset of the (full) OWS main collection~\cite{ows-book}, thus our samples of these two collections may overlap in parts, which could partially account for their extremely similar quality distributions.
Surprisingly, the Jensen-Shannon distance between the distributions of quality scores computed on OWS legal collection and ClueWeb22-B (en) is lower than the one between the two distributions computed on  OWS collections.
Notably, all three datasets also have very high quality overall. This is not surprising since ClueWeb22-B consists of the most frequently visited pages, which are thus highly relevant to users and of good quality~\cite{overwijk2022clueweb22}.
Similarly, the OWS datasets are pre-filtered for malicious URLs and spam documents using an exclusion list~\cite{hendriksen2024open} and thus should not include any extremely low-quality pages.
As already mentioned, since the two OWS datasets we consider neither provide a web graph, nor a query set for retrieval tasks associated with relevance labels, we conduct all subsequent experiments on ClueWeb22-B (en). However, since the quality distributions of OWS datasets and ClueWeb22-B (en) are very similar, our findings on ClueWeb22-B likely also hold true for OWS data.

\subsection{Relationship between quality \& relevance}
\begin{figure} 
    \centering
    \begin{minipage}[t]{0.25\textwidth}
        \centering
        \includegraphics[scale=0.55]{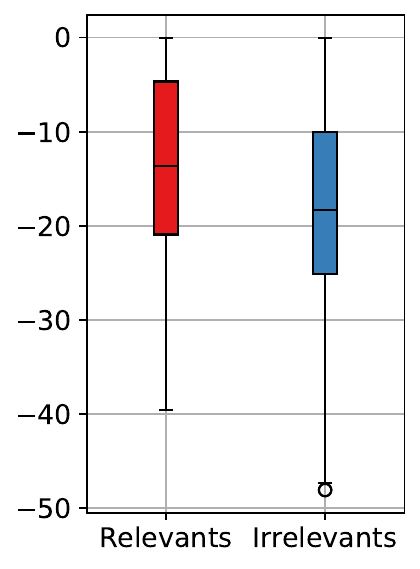}
    \end{minipage}
    \begin{minipage}[t]{0.7\textwidth}
        \centering
        \includegraphics[scale=0.55]{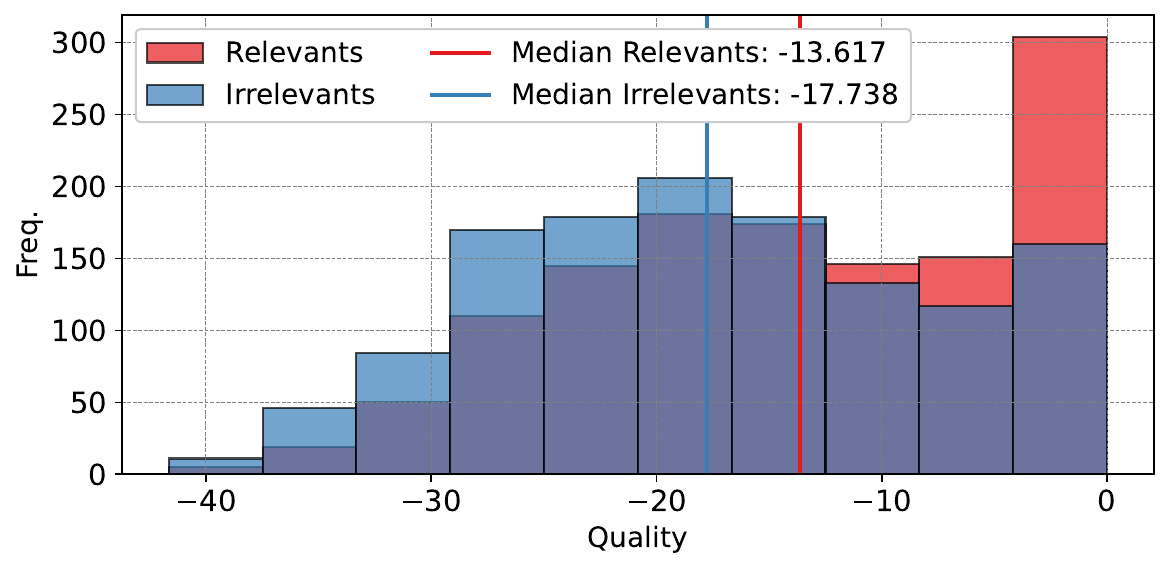}
    \end{minipage}
    \caption{Boxplot and histogram plot of the distribution of the quality scores of relevant and irrelevant web pages for $2867$ judged queries from the union of MSM-WS with RQ. However, in the histogram plot, to make the distributions comparable, we performed undersampling.}
    \label{fig:hists_qscore}
\end{figure}
On ClueWeb22-B (en), we first investigate whether quality scores provide good relevance signals, aiming to understand if by prioritising web pages with high semantic quality during the crawl, \textit{relevant} pages are automatically prioritised. To this end, we consider as \textit{relevant}, all the web pages of ClueWeb22-B (en) that have been judged relevant for at least one query belonging to the union of the entire RQ and MSM-WS query sets. The remaining documents are considered \textit{irrelevant}. In Figure~\ref{fig:hists_qscore} we show how the quality distributions  of relevant and irrelevant documents differ.
In particular, from Figure~\ref{fig:hists_qscore}, we note that relevant web pages generally exhibit higher quality than irrelevant ones, suggesting that quality scoring using neural estimators is a useful heuristic for distinguishing relevant pages from irrelevant ones.
Hence, the quality score is a promising relevance signal.
However, the large overlap between the two distributions indicates that the quality score alone does not allow for a perfect distinction. Thus, future work could focus on combining the relevance signal coming from neural quality scorers with other types of relevance signals to enhance effectiveness. Since quality scores are good relevance signals, we expect that employing quality scorers to prioritise high-quality pages during the crawling process could increase the likelihood of discovering relevant web pages early on.

\subsection{Downstream retrieval effectiveness of a \qoracle}
\begin{figure}
    \centering
    \includegraphics[width=0.7\linewidth]{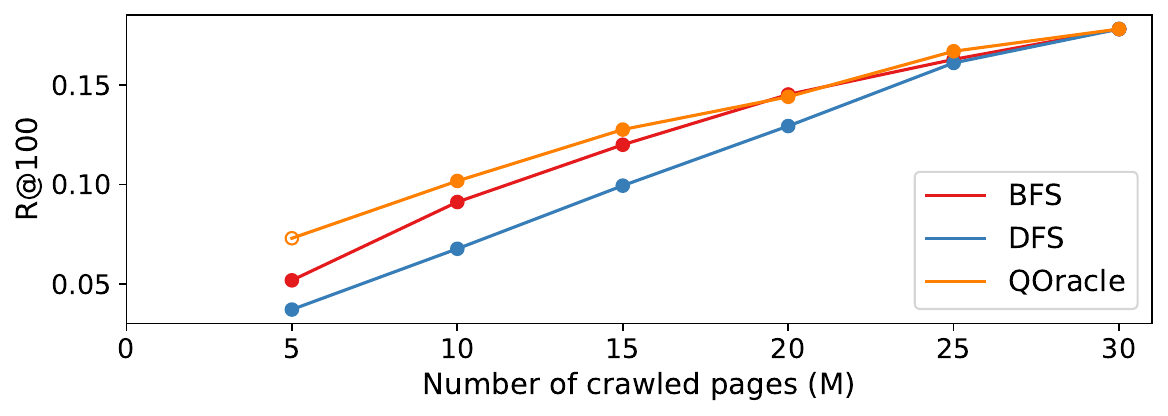}
    \caption{Comparison between the downstream retrieval effectiveness of our \qoracle\ crawler, and baseline BFS and DFS crawlers in terms of $R@100$ computed on a mixed query set composed of $850$ judged queries from RQ, and $850$ judged queries from MSM-WS. \textit{Hollow} markers denote statistically significant differences w.r.t. the two baselines, whereas \textit{filled} markers denote differences that are not statistically significant.}
    \label{fig:joined_recall}
\end{figure}
As a proof of concept, we next explore if prioritising semantically high-quality pages during the crawling process -- leveraging neural quality scorers -- can improve the downstream retrieval effectiveness of a search system. 
To investigate this, in Figure~\ref{fig:joined_recall} we show at different points in time, the retrieval effectiveness of a BM25 retriever measured with R@100 on search corpora built by our proposed oracle crawler (\qoracle). A Breadth-First-Search (BFS) crawler and a Depth-First-Search (DFS) crawler are shown as baselines. Our results demonstrate that the proposed \qoracle\ crawler outperforms both baselines in terms of early retrieval effectiveness. Therefore, guiding the crawl with a prioritisation strategy based on semantic quality scoring has a positive impact on downstream search effectiveness.

\subsection{Quality of outlinks}
\begin{figure}
    \centering
    \includegraphics[width=0.7\linewidth]{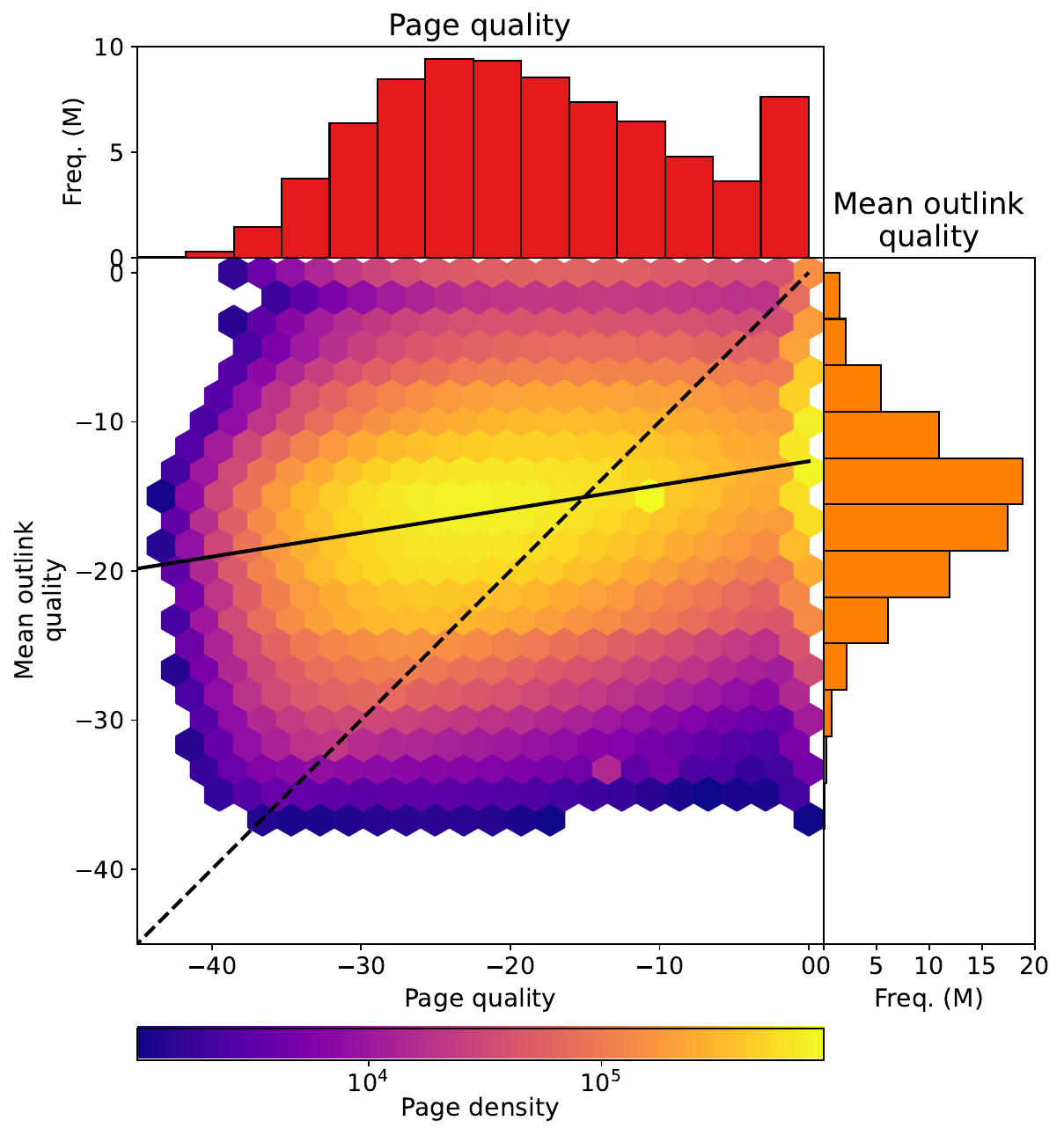}
    \caption{Visualisation of the weak positive linear correlation between the quality of a page ($x$-axis, histogram on top) and the average quality of the pages it links to ($y$-axis, histogram on right). Both the histograms are generated using $15$ bins. The hexagonal binning plot is created using a grid-size of $25$, and displays only hexagonal cells containing at at least $10^3$ pages. The \textit{solid} line represents the linear regression line of the data, while the \textit{dashed} line represents the theoretical linear regression line that the data under perfect positive correlation.}
    \label{fig:qual_links_qual}
\end{figure}
Finally, to gain an insight into the applicability of quality scoring in web crawling without an oracle function, we investigate if web pages mostly link to pages with similar semantic quality.
To address this, we filter ClueWeb22-B (en) for pages containing at least one outgoing link to another page in the same dataset. For pages in this subset, we compute the  the semantic quality of the page as well as the quality of the pages it links to. Next, we plot them in the hexagonal binning plot enriched by two marginal histograms, shown in Figure~\ref{fig:qual_links_qual}. 
This plot shows how the quality of a web page correlates with the average quality of the pages it links to. We note that there is a weak positive linear correlation between these two variables, also confirmed by the \textit{Pearson correlation coefficient} between the two, which is $0.286$. 

Additionally, we observe that most web pages have medium semantic quality and link to pages of similar quality. Meanwhile, web pages of very high semantic quality rarely link to web pages of extremely low quality. Therefore, when a crawler follows the outgoing links of high-quality pages, it is unlikely for it to discover very low-quality web pages. 
At the same time, very low-quality web pages generally do not link to high-quality pages.
As a result, low-quality pages can be de-prioritised without the risk of missing a significant number of high-quality pages.
These findings suggest that there may be a clear separation between very high-quality and very low-quality pages.
Consequently, crawlers can use this quality estimation to prioritise the crawling of high-quality pages, aiming to discover other valuable content while reducing the risk of wasting time and resources on low-quality pages.

\section{Conclusion \& Future Work}\label{conclusion}
In this paper, as part of our software contribution, we developed a Docker container that applies a neural quality scorer. This quality scoring module can be easily used and integrated in various components of web search systems to estimate the semantic quality of documents.
Additionally, we introduced an effective crawling approach that, by leveraging neural quality scorers, prioritises pages of high semantic quality. Our early experimental analysis performed on ClueWeb22-B (en), suggests that the prioritisation of semantically high-quality web pages during the crawl could effectively mitigate the negative impact of low-quality content on downstream retrieval effectiveness. Furthermore, we show that these findings are likely to generalise to Open Web Search datasets.
However, our findings also suggest that relevance signals coming from quality scoring should be combined with other signals for improved performance. This is a promising topic for future work, along with the exploration of crawling prioritisation strategies based on approximated quality scores.

 \begin{acknowledgments}
This work was partially supported by
the Spoke ``FutureHPC \& BigData'' of the ICSC – Centro Nazionale di Ricerca in High-Performance Computing, Big Data and Quantum Computing funded by the Italian Government, the FoReLab and CrossLab projects (Departments of Excellence), the NEREO PRIN project funded by the Italian Ministry of Education and Research and European Union - Next Generation EU (M4C1 CUP 2022AEF-HAZ), and the FUN project (SGA 2024FSTPC2PN30) funded by the OpenWebSearch.eu project (GA 101070014).
\end{acknowledgments}

\section*{Declaration on Generative AI}
The authors have not employed any Generative AI tools.
  
\bibliography{bibliography}

\end{document}